\documentclass{aa}  
\usepackage{graphicx}
\usepackage{txfonts}
\usepackage{lipsum}
\usepackage{subcaption}        
\usepackage[version=4]{mhchem}
\usepackage{lscape}
\usepackage{placeins}
\usepackage{natbib}
\bibpunct{(}{)}{;}{a}{}{,}
\usepackage{booktabs}
\usepackage{threeparttable}
\usepackage{siunitx}
\usepackage{stfloats}
\usepackage[colorlinks=true, linkcolor=blue, citecolor=blue, urlcolor=blue]{hyperref}

\begin{document}
\renewcommand{\linenumbers}{}
\renewcommand{\nolinenumbers}{}

\titlerunning{Beyond Idealized PAHs}
\authorrunning{Tan et al.} 

  \title{Beyond Idealized PAHs: Infrared Signatures of Carbon-Chain Defects from Shock Synthesis}

  \author{Xiaoting Tan \inst{1}
  \and Zhao Wang \inst{1}\fnmsep\thanks{Corresponding author: \email{zw@gxu.edu.cn}}
  \and Zu-Jia Lu \inst{1,2}
  \and Houda Haidar \inst{2}
  \and Dimitra Rigopoulou \inst{2,3}
}

   \institute{Laboratory for Relativistic Astrophysics, Department of Physics, Guangxi University, 530004 Nanning, China
   \and
   Department of Physics, University of Oxford, Keble Road, Oxford OX1 3RH, UK
   \and
   School of Sciences, European University Cyprus, Diogenes street, Engomi, 1516 Nicosia, Cyprus 
   }

\abstract
  {Polycyclic aromatic hydrocarbons (PAHs) are widely recognized as carriers of the aromatic infrared bands (AIBs). However, most spectral models rely on idealized structures that fail to capture the energetic environments of interstellar PAH formation.}
  {This work investigates the infrared (IR) signatures of PAHs formed under shock conditions and explores whether produced defective structures can explain observational features unpredicted by standard, idealized models.}
  {We combine two-stage reactive molecular dynamics simulations of PAH formation via condensation and shock processing with density functional theory spectral calculations, and compare our theoretical results with James Webb Space Telescope (JWST) observations of NGC~7023 and MRK~1066.}
  {Shock processing produces PAHs featuring fullerene-like carbon skeletons and linear carbon-chain attachments. These structural defects yield distinct IR signatures, including prominent carbon-chain stretching features at \SIrange{4.6}{5.5}{\micro\meter} that is absent in ideal PAHs, and significantly enhanced out-of-plane skeletal modes in the \SIrange{14.5}{20.0}{\micro\meter} regime.}
  {Our findings attribute the observed \SI{5.2}{\micro\meter} band in NGC~7023 and MRK~1066 to carbon-chain vibrations and the \SIrange{15}{18}{\micro\meter} emission to curved skeletal modes, providing observational support for the prevalence of defective, shock-formed PAHs in the interstellar medium.}

\keywords{ISM: molecules -- Infrared: ISM -- ISM: dust, extinction}
  
\maketitle

\section{Introduction}

Polycyclic aromatic hydrocarbons (PAHs) are major astrophysical carbonaceous species, widely accepted as carriers of the aromatic infrared bands (AIBs) at 3.3, 6.2, 7.7, 8.6, and \SI{11.2}{\micro\meter}, which arise from vibrational relaxation following UV absorption \citep{Allamandola1989, Leger1989}. Observed across diverse environments from circumstellar envelopes to early galaxies, their near-ubiquitous emission reflects their high abundance and resilience \citep{Tielens2008, Peeters2011}. Beyond serving as a key interstellar carbon reservoir, PAHs act as sensitive probes of astrophysical conditions \citep{Peeters2004, Galliano2008}. Because their IR spectra are dictated by molecular size, charge, and structure \citep{Bauschlicher2008}, decoding structure-spectral relationships is essential for extracting physical insight from astronomical observations.

The interpretation of astronomical PAH spectra has evolved significantly over recent decades \citep{Li2020, Peeters2021}. Benchmark dust models reproduced observed AIB features using continuous size distributions and empirical cross-sections \citep{Draine2001}. Subsequently, the NASA Ames PAH IR Spectroscopic Database (PAHdb) enabled structure-specific spectral fitting, providing quantitative constraints on PAH size, charge, and hydrogenation \citep{Boersma2014, Bauschlicher2018, Mattioda2020, Rigopoulou2021, Maragkoudakis2022, Kerkeni2022}. More recently, the high sensitivity and spatial resolution of the James Webb Space Telescope (JWST) have transformed PAH studies across local and high-redshift extragalactic environments \citep{Spilker2023, Rigopoulou2024}. In parallel, machine-learning frameworks have emerged as powerful tools to efficiently map the chemical space of PAHs and directly infer molecular properties from spectra \citep{Meng2023ML, Wang2026}. Collectively, these advances have greatly sharpened our understanding of space-borne PAHs.

Despite, existing studies share a fundamental limitation: they predominantly assume idealized PAH structures. Even when spectral databases include defective or functionalized species, modifications are typically limited to minor perturbations of a perfect benzenoid backbone, such as 5--7 ring pairs \citep{Ricca2011, Yu2012}, heteroatom substitutions \citep{Hudgins2005, Vats2022}, functional groups \citep{Yang2017, McGuire2021}, or vacancies \citep{Buragohain2018}. In essence, the molecular scaffold remains close to that of an ideal PAH. This narrow range of explored structural variations is unlikely to capture the diversity of interstellar PAHs, which chemically form under energetic conditions that yield intrinsically irregular, non-ideal architectures, possibly including non-fused rings, extended carbon chains, aliphatic sidegroups, irregular edges, and partial hydrogenation \citep{Parneix2017, Hanine2020, Meng2023MD, Patch2025}.

Interstellar PAHs form in dynamic, highly energetic environments, such as surface condensation on dust grains followed by processing in shocks, photodissociation regions (PDR), or protostellar outflows \citep{Barsony2010, Tappe2012, Montillaud2013}. Laboratory experiments and atomistic simulations demonstrate that shock processing of simple PAHs induces dehydrogenation, fragmentation, and re-assembly into defect-rich graphene-like nanostructures \citep{Weippert2020, Singh2025}. Because defective structures can display IR features markedly different from those of pristine PAHs, neglecting them introduces a critical blind spot: if interstellar PAHs are inherently irregular, properties inferred using perfect-molecule templates may be systematically biased.

This gap persists largely because dynamic formation pathways are computationally difficult to incorporate into spectral workflows via traditional quantum chemical methods \citep{Parker2015, Chen2020}, which scale steeply with system size. To bridge this gap, we here adopt a reactive molecular dynamics (MD) approach to simulate PAH synthesis under shock conditions, and compute the IR spectra of the \textit{in situ} formed molecules via density functional theory (DFT) calculations. By explicitly connecting chemical formation history to vibrational emission, this work evaluates how structural imperfections alter the spectral signatures of interstellar PAHs.

\section{Methods}

\subsection{Reaction Mechanism and Simulation Workflow}

We adopt a two-step formation mechanism for interstellar PAHs, as commonly invoked in the literature \citep{Cherchneff2011}. First, C and H atoms, along with small hydrocarbons such as CH or \ce{C2}, condense on dust or ice grain surfaces to form precursors \citep{Tsuge2023}. Second, these precursors undergo energetic processing, e.g. in protostellar outflows, where they are ejected into the hot gas and react to form PAHs \citep{Bachiller1996, Arce2007, Bossion2024}. The violent impact of a bipolar jet or outflow on the surrounding molecular cloud produces shocks, heating the aggregated material to several hundred or thousand K \citep{Draine1993}. Under such conditions, gas-phase chemistry efficiently converts the ejected precursors into PAH molecules. 

\begin{figure*}
\centerline{\includegraphics[width=0.9\textwidth]{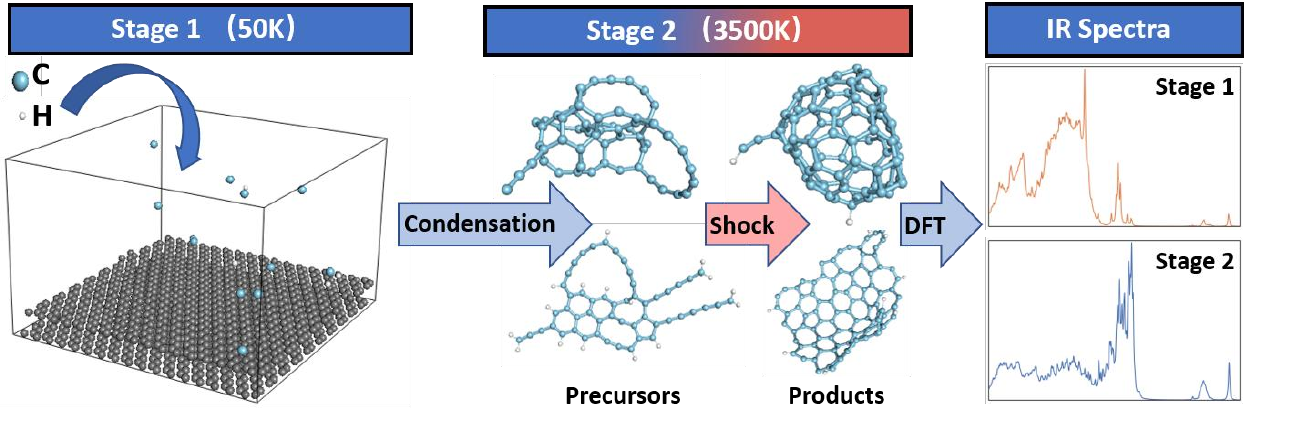}}
\caption{Schematic of the two-stage simulation workflow for PAH formation and spectral modeling. In Stage~1, C and H atoms condense onto a \SI{50}{\kelvin} graphene surface (with partial gas-phase reaction) to generate precursors. In Stage~2, the system is heated to \SI{3500}{\kelvin} to simulate shock processing, driving precursors to form larger, defective PAHs. The IR spectra of the final products are subsequently calculated using DFT.}
\label{F1}
\end{figure*}

Following this two-step mechanism, we design a two-stage MD simulation workflow (Figure~\ref{F1}). In Stage 1, we model the low-temperature condensation and adsorption of C and H atoms onto a simplified interstellar dust surface. This stage produces small hydrocarbon clusters and basic molecules that act as PAH precursors. In Stage 2, we apply high-temperature conditions to simulate shock processing of these precursors, driving the formation of larger PAHs and fullerene-like structures. Finally, we compute the IR spectra of the resulting molecules using DFT calculations and compare them with those of canonical, idealized PAHs. The details of each stage are described below.

\subsection{Stage 1: Atom Condensation onto Dust Surfaces}

We perform MD simulations to model the condensation of C and H atoms onto a dust surface. Although interstellar dust exhibits diverse compositions and morphologies, we adopt a graphene surface as a simplified proxy \citep{Marshall2016}. We note that this scenario is idealized and does not account for variations in grain properties. Graphene is chemically inert and thus captures physisorption effects, though it neglects chemisorption as a limitation acceptable for cold ISM environments. The graphene sheet is placed in a simulation box of dimensions $42.60 \times 41.81 \times 45.00$~\AA$^3$, with the $z$-axis normal to the surface, and periodic boundary conditions applied in the $x$ and $y$ directions to emulate an infinite surface.

C and H atoms are gradually deposited at random positions over \SI{5000}{\pico\second}, with $N_\mathrm{H}/N_\mathrm{C}$ ratios of 0.0, 0.1, 0.2, 0.4, 0.6, and 0.8. The deposited atoms are drawn to the surface and to each other by van der Waals interactions, where they react upon adsorption. Some atoms may also react in the gas phase before reaching the surface, forming small molecules such as \ce{C2} or \ce{CH}. Interatomic interactions are described by the adaptive interatomic reactive empirical bond order (AIREBO) force field \citep{Stuart2000}, which smoothly transitions between long-range and covalent interactions, enabling bond formation and breaking. The AIREBO model is well-established for $sp^2$-hydrocarbon systems and has been validated for diffusion and adsorption studies.

In our simulations, the system is equilibrated in the canonical ensemble (NVT), with the graphene sheet coupled to a thermostat at \SI{50}{\kelvin} using a Nos\'e-Hoover thermostat, while the deposited atoms evolve under Newtonian dynamics in the microcanonical ensemble (NVE). While typical ISM temperatures are \SIrange{10}{20}{\kelvin}, we adopt \SI{50}{\kelvin} to accelerate reactions as a standard practice in MD simulations. A time step of \SI{0.25}{\femto\second} is used. Atom deposition occurs on-the-fly over the first few nanoseconds, followed by a total run of \SI{5}{\nano\second}. A reflecting wall at $z = \SI{24.9}{\angstrom}$ prevents atoms from escaping.

\subsection{Stage 2: Shock Processing}

In the second stage, we simulate the evolution of the Stage 1 precursors under C-type shock conditions. A C-type shock dissipates energy gradually through ion-neutral friction over an extended region, heating the neutral gas more gently than J-type shock to peak temperatures of 1000-3000~K and allowing molecules to survive \citep{Skretas2025}. This warm reservoir promotes endothermic reactions, making C-type shocks particularly relevant for PAH synthesis, and we therefore focus on this regime.

The small molecules formed in Stage 1 desorb from the surface at temperatures above a few hundred Kelvin and are then heated to 1000-3000~K in the gas phase. To accelerate reactions as a standard MD practice for bridging short simulation timescales to astrophysical processes, we set the peak temperature to \SI{3500}{\kelvin}. The system is linearly ramped from \SI{50}{\kelvin} to \SI{3500}{\kelvin} over \SI{1.5}{\nano\second}, followed by a \SI{2.5}{\nano\second} isothermal hold at \SI{3500}{\kelvin}, allowing sufficient time for bond rearrangements, ring closure, and dehydrogenation to yield the final products. The system is evolved in the NVT ensemble with a Nosé-Hoover thermostat and a time step of \SI{0.1}{\femto\second} to resolve high-frequency vibrations and reactions. The total simulation runs for \SI{4}{\nano\second}, encompassing both the heating ramp and isothermal hold. Reaction progress is monitored via the potential energy to ensure completion within the simulation time.

To accurately capture complex bond-breaking and bond-forming processes such as ring closure and dehydrogenation, we employ the ReaxFF reactive force field \citep{Duin2001} in this stage, preferring it over AIREBO, which is primarily designed for $sp^2$-hybridized systems and may artificially bias the formation of planar structures. ReaxFF treats chemical reactions via bond-order-dependent potentials and has been extensively validated for hydrocarbon systems; further details of the force field parameterization and validation are given in \citet{Ashraf2017}.

The spatial and temporal scales of our MD simulations ($n \sim \SI{2.5e19}{\per\centi\meter\cubed}$, $t \sim \SI{4}{\nano\second}$) differ by many orders of magnitude from actual interstellar environments ($n \sim 10^{-2}$--$10^5\,\mathrm{cm^{-3}}$, $t \sim 10^3$--$10^6\,\mathrm{yr}$). Crucially, however, timescale and density are physically coupled through collision kinetics: drastically increasing the density accelerates bimolecular encounter rates, effectively compressing long astrophysical reaction timescales into the MD time window.

To quantitatively validate this scale-bridging strategy, we estimate the half-life ($t_{1/2}$) of carbon-chain cyclization using kinetic collision theory:

\begin{equation}
\label{Eq1}
t_{1/2} = \frac{\ln 2}{A n e^{-E_a / RT}},
\end{equation}
where $A = \SI{1e-10}{\centi\meter\cubed\per\second}$ is the gas-kinetic collision pre-factor, and $E_a = \SI{23}{\kilo\cal\per\mole}$ is a representative activation barrier for carbon-chain cyclization \citep{Marsh2000}.

\begin{figure}
\centerline{\includegraphics[width=0.5\textwidth]{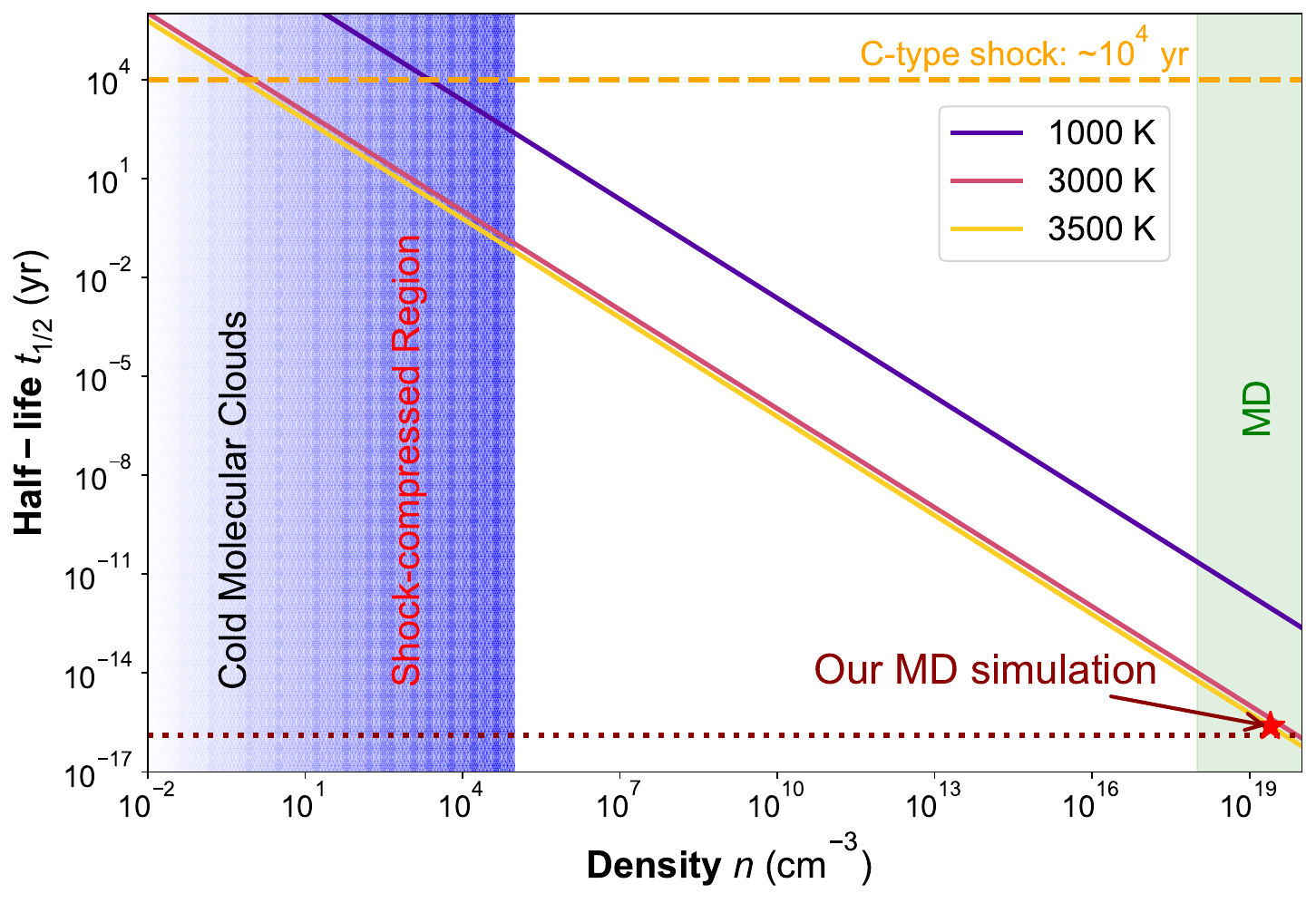}}
\caption{Half-life of PAH cyclization as a function of PAH number density for different temperatures, calculated from Eq.~\ref{Eq1}. The 1000 and 3000 K lines bracket the typical C-type shock temperature range, while the 3500 K line corresponds to our simulation temperature. Vertical bands indicate a region of PAH density transition from cold molecular clouds ($\sim 10^{-4}$--$10^{-1}$ cm$^{-3}$) to shock-compressed regions ($\sim 10^{2}$--$10^{5}$ cm$^{-3}$), and MD simulations ($\gtrsim 10^{18}$ cm$^{-3}$). The red star marks our MD condition ($n = 2.5\times 10^{19}$ cm$^{-3}$, $T = 3500$ K).}
\label{F2}
\end{figure}

As shown in Figure~\ref{F2}, $t_{1/2}$ scales inversely with density. Under realistic C-type shock conditions ($n \sim 10^1$--$10^4\,\mathrm{cm^{-3}}$, $T \sim 1000$--$3000\,\mathrm{K}$), the cyclization half-life spans $10^3$ to $10^6\,\mathrm{yr}$, comparable to or exceeding the shock duration. In contrast, at our MD density ($n \sim \SI{2.5e19}{\per\centi\meter\cubed}$) and peak temperature (\SI{3500}{\kelvin}), $t_{1/2}$ drops dramatically to $\sim 10^{-16}\,\mathrm{yr}$ ($\sim \SI{3}{\nano\second}$), well within our \SI{4}{\nano\second} simulation window. This kinetic equivalence demonstrates that density compression physically recovers the relevant chemical kinetics within achievable computational timescales.

We acknowledge that this high-density approach introduces potential caveats. Specifically, it suppresses slow radiative cooling and enhances kinetic trapping relative to low-density shocks, which may artificially inflate the yield of unclosed carbon chains and non-hexagonal defects. Nevertheless, because shock processing is intrinsically non-equilibrium, these kinetically trapped configurations represent physical, accessible local minima on the potential energy surface. Unlike equilibrium-based Monte Carlo methods that rely on pre-defined reaction paths, reactive MD uniquely preserves the explicit, non-equilibrium dynamics required to simulate shock-driven PAH formation.

\subsection{Quantum Chemical Calculation of IR Spectra}

To investigate the IR features of the resulting PAHs, we computed their harmonic IR spectra using DFT at the B3LYP/4-31G level, as implemented in Gaussian 16 \citep{Frisch2016}. This theoretical level offers a favorable balance between accuracy and computational efficiency for large PAHs and has been widely used in PAH IR studies. While the lack of polarization functions in 4-31G may introduce uncertainties for non-planar structures, previous benchmark studies show that B3LYP/4-31G with scaling factors reproduces major PAH vibrational features well \citep{Bauschlicher2000, Ricca2012}. Furthermore, test calculations on two curved defective PAHs using B3LYP/6-31G(d,p) yielded mean wavelength shifts of  \SIrange{0.104}{0.189}{\micro\meter} (post-scaling) and total integrated intensity differences of $0.53\%$--$3.68\%$ compared to 4-31G. Given the large size of our molecules ($N_{\mathrm{C}}$ up to 120), this level of theory remains a computationally practical and reliable choice for the present purpose.

Molecular geometries were optimized using the conjugate gradient algorithm, with negative frequencies carefully checked. Vibrational frequencies were computed under the double-harmonic approximation \citep{Bauschlicher2008}. To account for the curved geometries of our PAH products, long-range interactions were treated with Grimme's D3 empirical dispersion correction \citep{Grimme2010} and the Becke-Johnson damping function \citep{Grimme2011}. To correct for the systematic overestimation of bond strengths in DFT, we applied the frequency scaling factors recommended by \citet{Bauschlicher2018} for the 4-31G basis set: \num{0.956} for \SIrange{0}{1111.1}{\per\centi\meter}, \num{0.952} for \SIrange{1111.1}{2500}{\per\centi\meter}, and \num{0.960} above \SI{2500}{\per\centi\meter}.

The resulting vibrational modes and IR spectra were visualized and analyzed using GaussView \citep{Dennington2016}, which animates each normal mode as a displacement vector. This allows us to assign IR peaks to specific molecular motions (e.g., C--H stretching, ring breathing, or skeletal deformation). By linking each calculated frequency to its corresponding normal mode derived from the Hessian matrix of energy second derivatives with respect to nuclear displacements, we can identify the atomic motions responsible for each IR absorption band, thereby connecting spectral features to underlying structural motifs.

\section{Results and discussion}
\subsection{Formation and Structural Evolution}

\begin{figure}
\centerline{\includegraphics[width=0.5\textwidth]{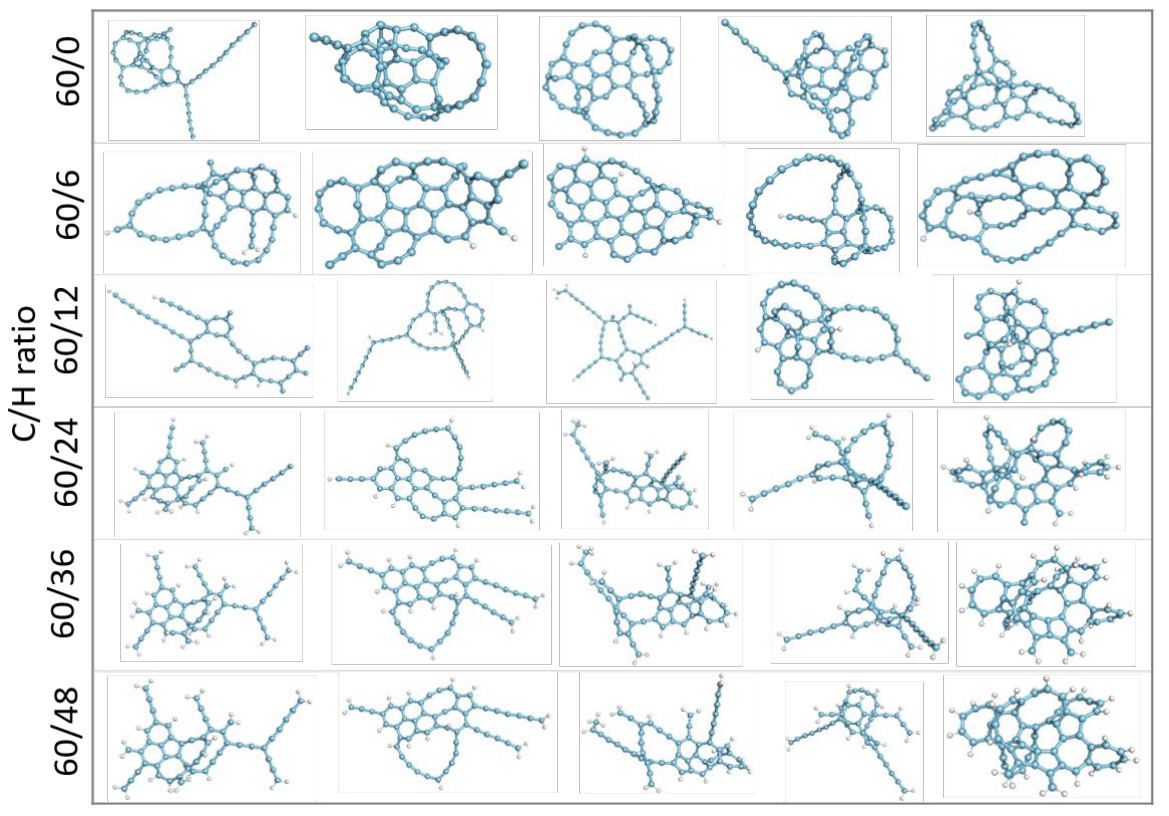}}
\caption{Snapshots of hydrocarbon precursors formed during the low-temperature condensation stage at different starting C/H ratios.}
\label{F3}
\end{figure}

We first examine the molecular structures of the precursors and products generated by our two-stage simulation workflow. Stage 1, which models low-temperature condensation of C and H atoms onto a dust surface, generates a diverse set of hydrocarbon clusters that serve as PAH precursors for subsequent shock-driven chemistry, as shown in Figure~\ref{F3} for different C/H ratios. The deposited atoms assemble into irregular, small PAH-like molecules containing carbon chains, pendant aliphatic groups, and a few $sp^3$-hybridized carbons. The low thermal energy limits surface diffusion and bond rearrangement, favoring kinetically trapped local configurations over thermodynamically preferred graphitic sheets. The C/H ratio strongly influences the precursor morphology: hydrogen-poor conditions promote larger, more extended rings by encouraging C--C bonding and ring closure, while hydrogen-rich conditions terminate dangling bonds with H atoms, favoring open carbon chains over closed rings. These disordered, non-planar aggregates are likely the types of precursors that, upon energetic processing such as shocks, may evolve into PAHs with structural defects \citep{Kroonblawd2019}.

\begin{figure}
\centerline{\includegraphics[width=0.5\textwidth]{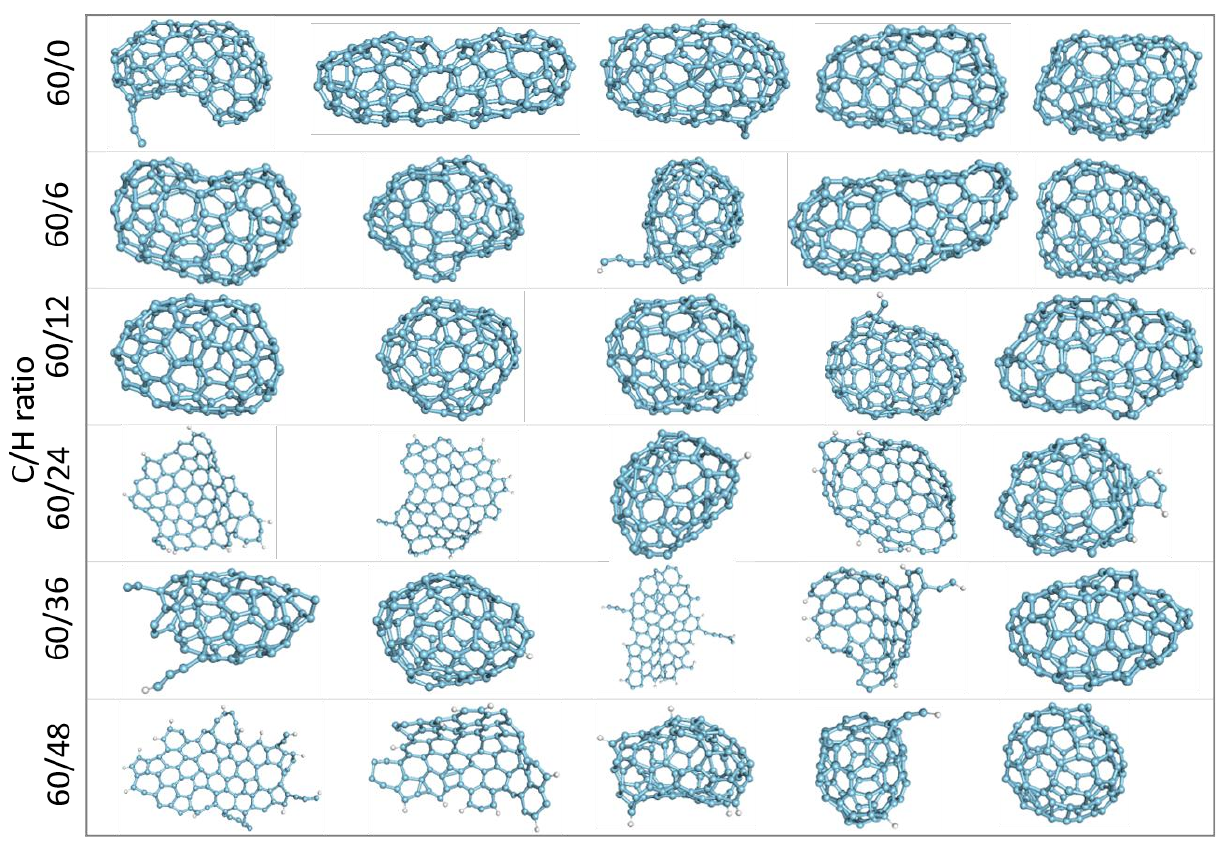}}
\caption{Representative products after the high-temperature shock stage for different starting C/H ratios (30 of 60 structures shown).}
\label{F4}
\end{figure}

We now turn to the transformation of these precursors under shock-like conditions in Stage 2. When heated to 3500 K, the precursors undergo dramatic gas-phase rearrangements. The thermal energy drives extensive bond breaking and reformation, promoting ring closure and dehydrogenation, ultimately yielding larger, more stable PAHs. Figure~\ref{F4} displays representative products obtained after the isothermal hold at 3500 K for different starting C/H ratios.

At high temperatures, most of the large rings present in the Stage 1 precursors disappear, giving way to curved, largely planar-extended PAHs with a pronounced three-dimensional warp \citep{Patra2014}. This is a consequence of the thermodynamic instability of strictly 2D networks in 3D space \citep{Mermin1968}. Some molecules resemble fullerene-like cages from certain orientations, but most remain unclosed, with openings or incomplete caps \citep{Meng2023MD}. Within the predominantly hexagonal carbon lattice, abundant pentagonal and heptagonal rings are interspersed, along with occasional short chain-like appendages at the edges. The prevalence of such 5-6-7 ring motifs indicates that shock-induced restructuring does not anneal the system to a perfect graphitic network; rather, the kinetics of bond rearrangement at 3500 K trap the system in metastable configurations containing topological defects, preventing full annealing to either planar graphene sheets or closed fullerenes.

The final morphology also slightly depends on the initial C/H ratio. Under hydrogen-poor conditions, extensive dehydrogenation drives the system toward fully closed, fullerene-like structures, as the lack of hydrogen allows the carbon network to curve and seal via pentagon formation. Conversely, under hydrogen-rich conditions, residual H atoms passivate reactive edge sites and prevent complete ring closure, resulting in open, curved PAHs that remain unclosed even after prolonged annealing. This trend mirrors the well-known preference for fullerene formation in hydrogen-poor environments and PAH dominance when hydrogen is abundant \citep{Cami2010}. These imperfect, warped structures, neither fully planar PAHs nor completely closed fullerenes, represent an intermediate morphological regime that has received little attention in astronomical spectral modeling.

\subsection{IR Spectral Signatures}

\begin{figure}
\centerline{\includegraphics[width=0.5\textwidth]{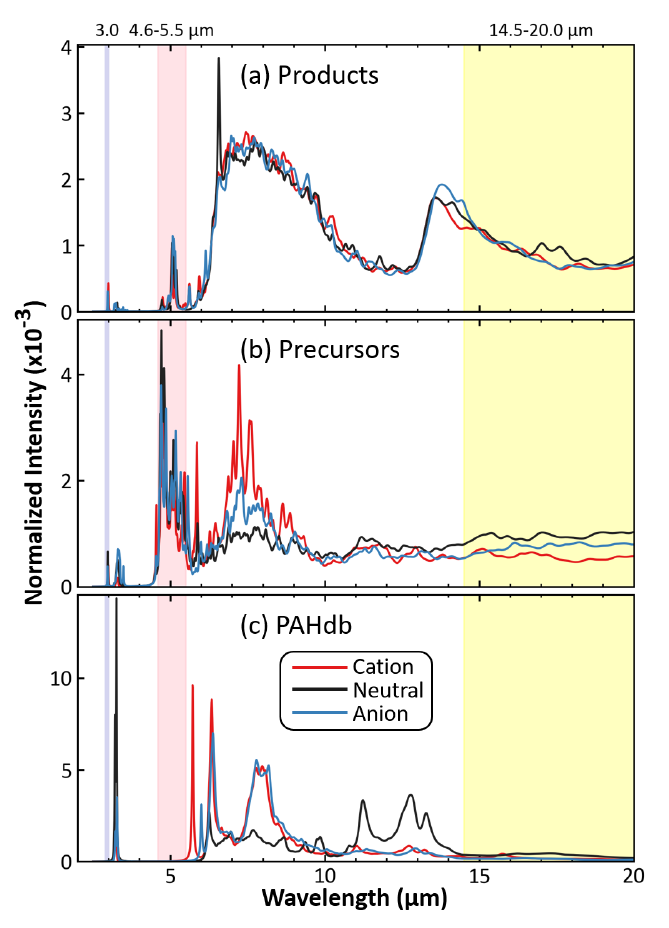}}
\caption{Stacked harmonic IR spectra for (a) 48 shock-processed products, (b) 41 precursors, and (c) the 735 hydrocarbon PAH structures from the NASA Ames PAHdb v4.0 (mean $N_{\mathrm{C}} \sim 77$). All spectra are normalized to unit integrated area.}
\label{F5}
\end{figure}

To assess how the structurally irregular PAHs formed in our two-stage simulations differ spectroscopically from their idealized counterparts, we computed the IR spectra of both the precursors (Stage~1) and products (Stage~2) via DFT calculations, for three charge states: neutral, anion, and cation. Figure~\ref{F5} presents the stacked IR spectra for the product (panel a) and precursor (panel b) structures, alongside the reference spectra from the NASA Ames PAHdb v4.0 \citep{Ricca2026}, which comprise 2,205 individual spectra for 735 hydrocarbon PAH structures with data available for all three charge states.

A direct comparison between the \textit{in situ} formed species (panel a) and the PAHdb reference (panel c) reveals that the spectra of our products exhibit a markedly weaker dependence on charge state than those of standard PAHdb molecules. Rather than being purely a size effect, given that the sizes are comparable (mean $N_{\mathrm{C}} \sim 84$ versus $77$ in PAHdb), this relative charge insensitivity may stem primarily from the 3D non-planar geometries of the shock-synthesized species. First, the 3D curved skeletons facilitate isotropic $\pi$-electron delocalization and suppress the charge-induced vibronic coupling (Jahn-Teller-like distortions) that typically drives the dramatic enhancement of the C--C stretching modes in ionized planar PAHs. Second, the intrinsically broken spatial symmetry in these curved and defective geometries creates large intrinsic dipole moments even in the neutral state. Consequently, the neutral-state skeletal modes already possess high intrinsic IR activity, rendering subsequent charge perturbations far less pronounced than in highly symmetric, planar PAHs.

Despite the overall spectral envelope broadly resembling that of canonical PAHdb PAHs, with prominent features near the C--H out-of-plane bending, C--C stretching, and C--H stretching regions \citep{Bauschlicher2009}, substantial discrepancies exist in both band positions and profiles. Three spectral regions, highlighted by the shaded spectral regions in Figure~\ref{F5}, stand out as particularly diagnostic of the irregular, non-planar nature of our products: the \SIrange{14.5}{20.0}{\micro\meter}, \SIrange{4.6}{5.5}{\micro\meter}, and \SI{3.0}{\micro\meter} regions. We examine each in turn below with the aid of normal-mode analyses performed using GaussView.

\begin{figure}
\centerline{\includegraphics[width=0.5\textwidth]{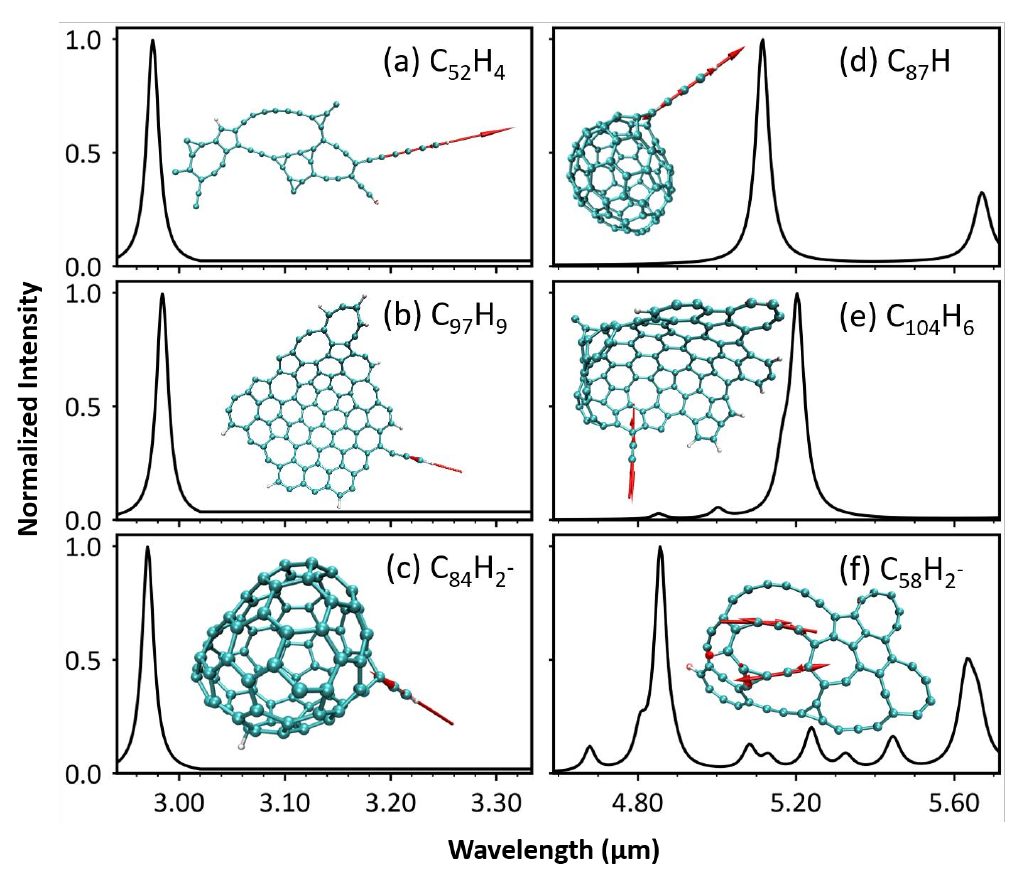}}
\caption{Localized spectral regions showing the diagnostic features of defective PAHs. Panels (a)--(c) cover the \SIrange{3000}{3400}{\per\centi\meter} range ($\sim$\SI{3.0}{\micro\meter}), and panels (d)--(f) cover the \SIrange{1750}{2180}{\per\centi\meter} range ($\sim$\SI{5}{\micro\meter}). The vertical axis denotes normalized intensity.}
\label{F6}
\end{figure}

The \SIrange{4.6}{5.5}{\micro\meter} region is particularly diagnostic, as canonical PAHs exhibit virtually no IR activity in this window, yet the \textit{in situ} formed species display distinct bands with non-negligible intensities. As shown in Figure~\ref{F6}d--f, normal-mode analysis attributes these features to stretching vibrations of linear carbon chains attached to the PAH periphery. Laboratory studies have attributed polyyne (\(-\mathrm{C}{\equiv}\mathrm{C}-\)) and cumulenic (\(=\mathrm{C}{=}\mathrm{C}{=}\)) chains to features near \SI{4.76}{\micro\meter} and in the \SIrange{5.08}{5.26}{\micro\meter} range, respectively \citep{Duley2009}. 

We note, however, that interstellar observations in this window may be complicated by CO ice (\SI{4.68}{\micro\meter}) and OCN$^-$ (\SI{4.62}{\micro\meter}) absorptions \citep{Pendleton1999, Duley2009}, as well as PAH overtones \citep{Boersma2009, Mai2025}. The persistence of these bands in our shock products confirms that such \textit{sp}-chains remain kinetically trapped even after high-temperature annealing. The appearance of $\sim$\SI{5}{\micro\meter} emission thus serves as a potential diagnostic for the presence of such defective, incompletely annealed carbon networks in interstellar environments, as a signature that would be entirely missed if one relied solely on perfect-molecule templates.

The \SI{3.0}{\micro\meter} region corresponds to stretching vibrations of C--H bonds at the termini of linear carbon chains attached to the PAH skeleton, as shown in Figure~\ref{F6}a--c. The persistence of these terminal C--H signatures after high-temperature annealing indicates that our shock-processed PAHs have undergone complex rearrangements that preserve linear carbon chains terminating in C--H bonds, rather than simply dehydrogenating into compact aromatic networks. Observationally, the presence or absence of $\sim$\SI{3}{\micro\meter} features from such chain termini could provide constraints on the formation history and processing conditions of interstellar PAH populations, although their intrinsically weak intensity may make them challenging to detect against the stronger aromatic C--H stretching bands near \SI{3.3}{\micro\meter}.

\begin{figure}
\centerline{\includegraphics[width=0.5\textwidth]{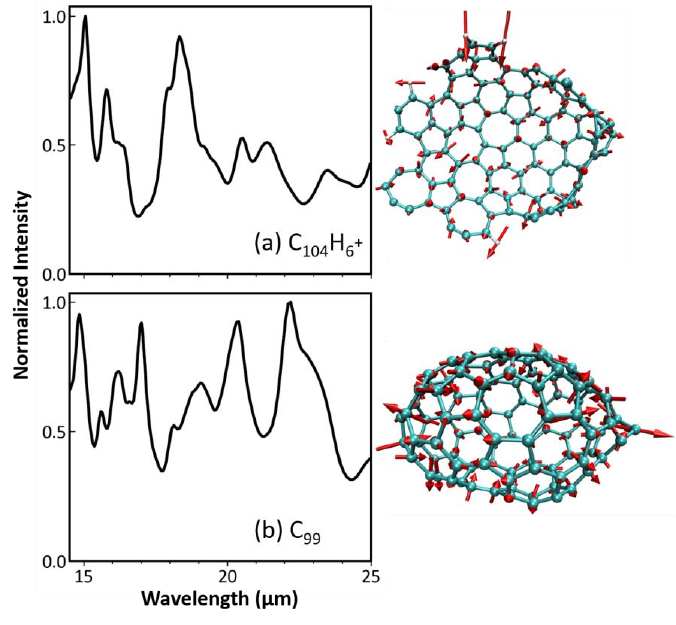}}
\caption{Localized spectral region (\SIrange{400}{700}{\per\centi\meter}, $\sim$\SIrange{14.5}{25}{\micro\meter}) and corresponding vibrational mode assignments. The arrows indicate atomic displacement directions.}
\label{F7}
\end{figure}

We now turn to the broader \SIrange{14.5}{20}{\micro\meter} region (Figure~\ref{F7}), which encompasses out-of-plane vibrational modes of the carbon skeleton that are sensitive to the overall planarity of the molecular geometry. In the PAHdb reference (Figure~\ref{F5}c), this region typically exhibits only very weak features, as the out-of-plane modes of planar, symmetric PAHs are largely IR-inactive. In stark contrast, our \textit{in situ} formed products display a significantly stronger and broader spectral feature across this entire window. This dramatic enhancement is attributable to the curved, fullerene-like character of our molecules: the non-planar carbon network allows out-of-plane skeletal deformations to induce significant changes in the molecular dipole moment, rendering these modes highly IR-active. Figure~\ref{F7} presents the localized spectra in the \SIrange{14.5}{20.0}{\micro\meter} range alongside the normal-mode assignments, which confirm that the observed bands originate from out-of-plane puckering and deformation of the carbon skeleton.

This behavior parallels the IR characteristics of interstellar fullerenes. Neutral \ce{C60}, with its icosahedral symmetry, possesses only four IR-active modes, located at  7.0, 8.5, 17.4, and 18.9~\si{\micro\meter} \citep{Cami2010}. The 17.4 and 18.9 \si{\micro\meter} bands fall precisely within our diagnostic window and arise from out-of-plane cage deformations. \citet{Sellgren2010} showed that NGC 7023 and NGC 2023 exhibit IR emission in the \SIrange{14.5}{20}{\micro\meter} region, attributed to fullerene carriers. Our results suggest that the enhanced \SIrange{14.5}{20}{\micro\meter} emission from shock-synthesized PAHs shares a common physical origin with fullerene bands: curvature-induced activation of out-of-plane skeletal modes.

\section{Observational Implications}

The spectroscopic differences identified above suggest that derived physical parameters, e.g. ionization fraction, size distribution, and UV field intensity, may be subject to systematic biases if a substantial fraction of interstellar PAHs resemble our curved, defective products rather than the idealized structures. Our results point to several diagnostic band ratios for identifying defective PAHs in space. The $\sim$\SI{5.2}{\micro\meter} feature traces carbon-chain content, while the \SIrange{14.5}{20}{\micro\meter} serves as a tracer of skeletal curvature, as out-of-plane skeletal modes are substantially enhanced in our fullerene-like products. The $\sim$\SI{5.2}{\micro\meter} feature is particularly promising as a diagnostic because it arises from stretching vibrations of linear carbon chains, which are entirely absent in canonical, idealized PAHs.

\citet{Boersma2009} assigned the 5.25 and 5.7~\si{\micro\meter} features to CH overtone/combination bands, but this interpretation faces a critical intensity problem: overtones are intrinsically weak, relying on anharmonic borrowing from fundamentals, yet the observed relative intensity of the 5.25~\si{\micro\meter} feature is remarkably high for a second-order process. Indeed, \citet{Boersma2009} acknowledged that their calculations could not determine the intensities of this band, leaving the strength unexplained. Our carbon-chain model offers an alternative interpretation: the emission arises from fundamental stretching modes of linear carbon chains, which are inherently strong and require no anharmonic borrowing. 

A potential constraint on carbon-chain carriers is their photochemical stability under intense interstellar UV fields. While bare $sp$-chains readily photodissociate, attachment to a large PAH scaffold enables redistribution of absorbed UV energy across the aromatic bulk via internal conversion, enhancing chain survival. Furthermore, while this work focuses on shock synthesis, carbon-chain-bearing PAHs in PDRs like NGC 7023 may also originate from the partial photolytic destruction and fragmentation of larger PAHs; together with ongoing shock processing and grain resupply, these mechanisms maintain a steady-state population of carbon-chain carriers against complete degradation.

\begin{figure*}
\centerline{\includegraphics[width=0.95\textwidth]{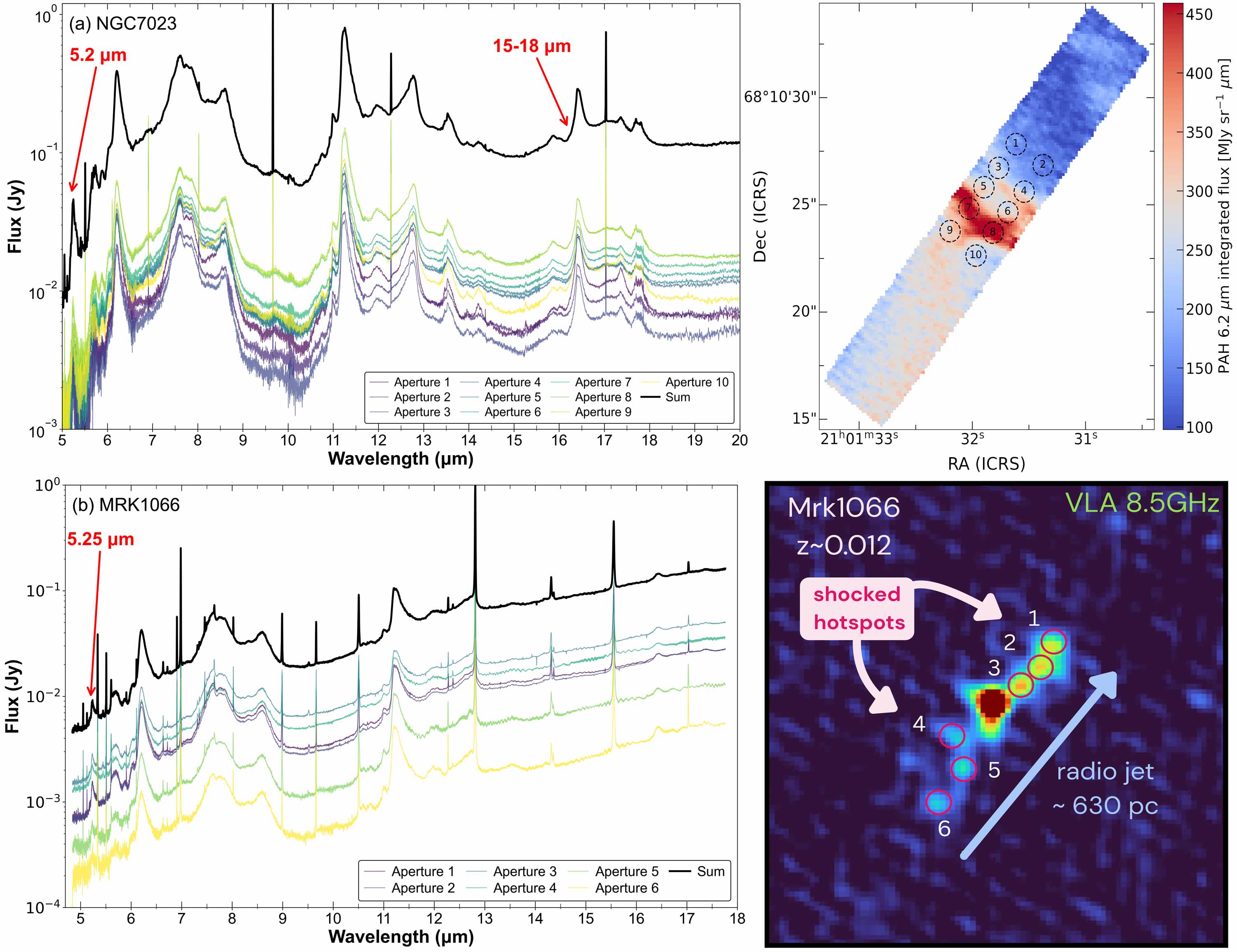}}
\caption{(a) JWST MIRI-MRS spectra of NGC~7023. \textit{Left:} spectra of 10 circular apertures with a radius of 0.5 arcsec. The black spectrum is the sum of all 10 apertures. The colour coding of the 10 individual spectra traces the spatial variation across the PDR. \textit{Right:} map of the PAH \SI{6.2}{\micro\meter} integrated flux with the 10 aperture positions overlaid as black dashed circles.  (b) JWST MIRI-MRS spectra of MRK~1066. \textit{Left:} spectra of six circular apertures with a radius of 0.2 arcsec.  \textit{Right:} VLA radio (8.5 GHz) map showcasing the radio jet in Mrk 1066 and several hot spots over which the apertures were selected. }
\label{F8}
\end{figure*}

Features around \SI{5}{\micro\meter} have been observed in a variety of astrophysical sources, including H\,II regions, reflection nebulae, planetary nebulae, post-asymptotic giant branch (post-AGB) stars, and Herbig Ae/Be stars, with the Short Wavelength Spectrometer (SWS) on board the Infrared Space Observatory (ISO) \citep{Boersma2009, Peeters2002, vanDishoeck2004}. Here we present more recent observations with the Mid-Infrared Instrument-Medium Resolution Spectrometer (MIRI-MRS) on board JWST, focusing on the NGC~7023 (JWST Program ID: 1192, PI: K. Misselt) \citep{Misselt2025A&A} and the Seyfert~2 galaxy Mrk~1066 (JWST Program ID: 7802, PI: H. Haidar), as shown in Figure \ref{F8}. 

The data reduction for both sources followed the standard JWST Science Calibration Pipeline (version 1.14.0). The MIRI-MRS data were processed with additional steps to correct for instrumental artifacts, including custom pixel flagging to identify bad pixels and image-to-image background subtraction. Details of the pipeline are given in \citet{Misselt2025A&A}. The resulting data cubes were then resampled onto a common spatial grid using the \texttt{reproject} package \citep{Robitaille2020}, and spectra were extracted from circular apertures with a radius of 0.5 arcsec at 10 distinct positions across the PDR front. 

The JWST/MIRI-MRS observations of NGC~7023 (Figure~\ref{F8}a) reveal two spectral features that are particularly relevant to our defect-PAH model. We detect a distinct emission peak near \SI{5.2}{\micro\meter}, with a relatively narrow profile and non-negligible intensity. This feature is difficult to attribute to CH overtone or combination bands: such second-order processes are not only intrinsically weak but also produce broad, smooth spectral profiles due to the involvement of numerous combination and difference frequencies, whereas the observed peak is comparatively sharp. The relatively narrow profile and strength of the observed peak are qualitatively consistent with fundamental stretching modes of linear carbon chains attached to PAH peripheries, as seen in our \textit{in situ} shock-synthesis products.

In the \SIrange{15}{18}{\micro\meter} region, we observe a prominent, broad emission feature that persists across all apertures in Figure~\ref{F8}a, which forms part of the well-known \SI{17}{\micro\meter} complex containing resolved sub-components at \SI{16.4}{\micro\meter} and \SI{17.4}{\micro\meter} \citep{Werner2004, Smith2007}. While the \SI{17.4}{\micro\meter} feature is commonly attributed to neutral $C_{60}$ fullerenes \citep{Cami2010, Sellgren2010}, the detailed carrier of the \SI{16.4}{\micro\meter} sub-feature and the underlying continuum remain debated. This emission falls within the \SIrange{14.5}{20}{\micro\meter} diagnostic window identified in our simulations for out-of-plane skeletal deformations, where such modes become IR-active only when the carbon skeleton acquires sufficient curvature to induce significant changes in the molecular dipole moment. The detection of this broad feature in NGC~7023 thus provides an attractive alternative explanation, suggesting that curved, defect-rich carbon networks, formed via shock processing or photolytic degradation, naturally contribute to the \SIrange{15}{18}{\micro\meter} envelope through enhanced skeletal puckering modes alongside fullerene carriers.

The detection of these features across multiple apertures indicates that the carriers are spatially extended within the PDR. NGC~7023 is a reflection nebula illuminated by HD~200775, which drives a prominent PDR \citep{Cesarsky1996}. Although PDRs and shocks are distinct regimes, they can coexist: the expanding molecular shell driven by the central star can generate weak shocks at the PDR interface \citep{Berne2013, VanDePutte2026}, and PAH emission in PDRs can be influenced by shock processing \citep{Rapacioli2005, Castellanos2014}. The northwest PDR has been extensively studied as a benchmark for carbonaceous grain evolution, with Spitzer observations detecting both PAH features and \ce{C60} fullerene emission in the central cavity \citep{Sellgren2010}. Recent JWST studies reveal at least two distinct spatial components in the 16--18~\si{\micro\meter} emission, suggesting multiple carrier populations \citep{VanDePutte2026}. Our detection of enhanced emission in this range, together with the $\sim$\SI{5.2}{\micro\meter} carbon-chain feature, supports the presence of structurally defective, curved species, likely shaped by UV photoprocessing and shock history, that may represent an intermediate stage in the photochemical evolution of PAHs toward fullerenes.

The emission feature near \SI{5.25}{\micro\meter} is also detected in the JWST/MIRI-MRS spectrum of MRK~1066, as presented in Figure~\ref{F8}b. The selected 6 apertures are aligned along the direction of the radio jet and outflow, where energetic shocks driven by the active galactic nucleus (AGN) are active. Unlike NGC~7023, where the broad emission feature between \SIrange{15}{20}{\micro\meter} is highly prominent, MRK~1066 displays a comparatively weaker continuum in this longer wavelength regime. This variation likely reflects the much harsher radiation field and strong dynamic interactions near the AGN core, where larger, highly curved fullerene-like structures may be more efficiently destroyed or ionized. 
Consequently, while the detection of the \SI{5.25}{\micro\meter} feature along the outflow suggests the plausible action of shock synthesis, a detailed analysis of PAH processing in shocks for this target will be presented in a follow-up paper (Haidar et al., in prep.).
Future spatially resolved observations combining millimeter and radio wavelengths (e.g., ALMA/VLA tracing dense shock tracers such as SiO) will be crucial to 
constraining the chemical destruction and reformation pathways of carbonaceous species in such extragalactic outflow environments.

\section{Conclusions}

Our two-stage reactive MD simulations demonstrate that shock-driven processing of hydrocarbon precursors yields PAHs with abundant topological defects, including curved non-planar carbon skeletons, and linear carbon chains, that remain kinetically trapped even after high-temperature annealing. The IR spectra of these structurally irregular PAHs exhibit distinct diagnostic features that are absent or negligible in idealised PAH templates, most notably in the \SIrange{4.6}{5.5}{\micro\meter} and \SIrange{14.5}{20.0}{\micro\meter} regions, which arise from carbon-chain stretching vibrations and out-of-plane skeletal deformations, respectively.

Our models provide a compelling origin for the interstellar \SI{5.2}{\micro\meter} and \SIrange{15}{18}{\micro\meter} emission features, which standard, canonical PAHs fail to explain. Based on JWST/MIRI-MRS observations of NGC~7023 and MRK~1066, we attribute the \SI{5.2}{\micro\meter} band to fundamental carbon-chain vibrations and the \SIrange{15}{18}{\micro\meter} excess to out-of-plane deformations of curved, fullerene-like structures. These signatures strongly point to a population of defective, shock-formed PAHs in dynamic interstellar environments. While these benchmark sources support this mechanism, a broader JWST spectroscopic survey across diverse PDRs, planetary nebulae, and galactic outflows is essential to robustly constrain the prevalence of such defective PAHs and decouple their emissions within the \SI{17}{\micro\meter} complex.

\section*{Data Availability}
The IR spectra and optimized geometries of the shock-processed PAHs generated in this work are publicly available on Zenodo at \href{https://doi.org/10.5281/zenodo.21507231}{DOI: 10.5281/zenodo.21507231}.

\begin{acknowledgements}
The authors acknowledge financial support from the National Natural Science Foundation of China (Grant Nos.~12463005, 12563005, and 12494571), the Guangxi Natural Science Foundation (Grant No.~2026GXNSFHA00640301), the National Key R\&D Program of China (Grant Nos.~2024YFA1611704 and 2024YFA1611700), and the Guangxi Talent Programme (Highland of Innovation Talents). DR and HH acknowledge the support from a Leverhulme Trust Research Project Grant.
\end{acknowledgements}


\begin{thebibliography}{78}
\expandafter\ifx\csname natexlab\endcsname\relax\def\natexlab#1{#1}\fi

\bibitem[{{Allamandola} {et~al.}(1989){Allamandola}, {Tielens}, \& {Barker}}]{Allamandola1989}
{Allamandola}, L.~J., {Tielens}, A.~G.~G.~M., \& {Barker}, J.~R. 1989, \apjs, 71, 733

\bibitem[{{Arce} {et~al.}(2007){Arce}, {Shepherd}, {Gueth}, {Lee}, {Bachiller}, {Rosen}, \& {Beuther}}]{Arce2007}
{Arce}, H.~G., {Shepherd}, D., {Gueth}, F., {et~al.} 2007, in Protostars and Planets V, ed. B.~{Reipurth}, D.~{Jewitt}, \& K.~{Keil}, 245

\bibitem[{{Ashraf} \& {van Duin}(2017)}]{Ashraf2017}
{Ashraf}, C. \& {van Duin}, A. C.~T. 2017, J. Phys. Chem. A, 121, 1051

\bibitem[{{Bachiller}(1996)}]{Bachiller1996}
{Bachiller}, R. 1996, \araa, 34, 111

\bibitem[{{Barsony} {et~al.}(2010){Barsony}, {Wolf-Chase}, {Ciardi}, \& {O'Linger}}]{Barsony2010}
{Barsony}, M., {Wolf-Chase}, G.~A., {Ciardi}, D.~R., \& {O'Linger}, J. 2010, \apj, 720, 64

\bibitem[{{Bauschlicher} \& {Bakes}(2000)}]{Bauschlicher2000}
{Bauschlicher}, C.~W. \& {Bakes}, E.~L.~O. 2000, Chem. Phys., 262, 285

\bibitem[{{Bauschlicher} {et~al.}(2008){Bauschlicher}, {Peeters}, \& {Allamandola}}]{Bauschlicher2008}
{Bauschlicher}, Jr., C.~W., {Peeters}, E., \& {Allamandola}, L.~J. 2008, \apj, 678, 316

\bibitem[{{Bauschlicher} {et~al.}(2009){Bauschlicher}, {Peeters}, \& {Allamandola}}]{Bauschlicher2009}
{Bauschlicher}, Jr., C.~W., {Peeters}, E., \& {Allamandola}, L.~J. 2009, \apj, 697, 311

\bibitem[{{Bauschlicher} {et~al.}(2018){Bauschlicher}, {Ricca}, {Boersma}, \& {Allamandola}}]{Bauschlicher2018}
{Bauschlicher}, Jr., C.~W., {Ricca}, A., {Boersma}, C., \& {Allamandola}, L.~J. 2018, \apjs, 234, 32

\bibitem[{{Bern{\'e}} {et~al.}(2013){Bern{\'e}}, {Mulas}, \& {Joblin}}]{Berne2013}
{Bern{\'e}}, O., {Mulas}, G., \& {Joblin}, C. 2013, \aap, 550, L4

\bibitem[{{Boersma} {et~al.}(2014){Boersma}, {Bauschlicher}, {Ricca}, {Mattioda}, {Cami}, {Peeters}, {S{\'a}nchez de Armas}, {Puerta Saborido}, {Hudgins}, \& {Allamandola}}]{Boersma2014}
{Boersma}, C., {Bauschlicher}, Jr., C.~W., {Ricca}, A., {et~al.} 2014, \apjs, 211, 8

\bibitem[{{Boersma} {et~al.}(2009){Boersma}, {Mattioda}, {Bauschlicher}, {Peeters}, {Tielens}, \& {Allamandola}}]{Boersma2009}
{Boersma}, C., {Mattioda}, A.~L., {Bauschlicher}, Jr., C.~W., {et~al.} 2009, \apj, 690, 1208

\bibitem[{{Bossion} {et~al.}(2024){Bossion}, {Sarangi}, {Aalto}, {Esmerian}, {Hashemi}, {Knudsen}, {Vlemmings}, \& {Nyman}}]{Bossion2024}
{Bossion}, D., {Sarangi}, A., {Aalto}, S., {et~al.} 2024, \aap, 692, A249

\bibitem[{{Buragohain} {et~al.}(2018){Buragohain}, {Pathak}, {Sarre}, \& {Gour}}]{Buragohain2018}
{Buragohain}, M., {Pathak}, A., {Sarre}, P., \& {Gour}, N.~K. 2018, \mnras, 474, 4594

\bibitem[{{Cami} {et~al.}(2010){Cami}, {Bernard-Salas}, {Peeters}, \& {Malek}}]{Cami2010}
{Cami}, J., {Bernard-Salas}, J., {Peeters}, E., \& {Malek}, S.~E. 2010, Science, 329, 1180

\bibitem[{{Castellanos} {et~al.}(2014){Castellanos}, {Bern{\'e}}, {Sheffer}, {Wolfire}, \& {Tielens}}]{Castellanos2014}
{Castellanos}, P., {Bern{\'e}}, O., {Sheffer}, Y., {Wolfire}, M.~G., \& {Tielens}, A. G.~G.~M. 2014, \apj, 794, 83

\bibitem[{{Cesarsky} {et~al.}(1996){Cesarsky}, {Lequeux}, {Abergel}, {Perault}, {Palazzi}, {Madden}, \& {Tran}}]{Cesarsky1996}
{Cesarsky}, D., {Lequeux}, J., {Abergel}, A., {et~al.} 1996, \aap, 315, L305

\bibitem[{{Chen} {et~al.}(2020){Chen}, {Luo}, \& {Li}}]{Chen2020}
{Chen}, T., {Luo}, Y., \& {Li}, A. 2020, \aap, 633, A103

\bibitem[{{Cherchneff}(2011)}]{Cherchneff2011}
{Cherchneff}, I. 2011, in EAS Publications Series, Vol.~46, EAS Publications Series, ed. C.~{Joblin} \& A.~G.~G.~M. {Tielens} (EDP), 177--189

\bibitem[{Dennington {et~al.}(2016)Dennington, Keith, \& Millam}]{Dennington2016}
Dennington, R., Keith, T.~A., \& Millam, J.~M. 2016, GaussView, Version 6, Semichem Inc., Shawnee Mission, KS, computer software

\bibitem[{{Draine} \& {Li}(2001)}]{Draine2001}
{Draine}, B.~T. \& {Li}, A. 2001, \apj, 551, 807

\bibitem[{{Draine} \& {McKee}(1993)}]{Draine1993}
{Draine}, B.~T. \& {McKee}, C.~F. 1993, \araa, 31, 373

\bibitem[{{Duley} \& {Hu}(2009)}]{Duley2009}
{Duley}, W.~W. \& {Hu}, A. 2009, \apj, 698, 808

\bibitem[{Frisch {et~al.}(2016)Frisch, Trucks, Schlegel, Scuseria, Robb, Cheeseman, Scalmani, Barone, Petersson, Nakatsuji, Li, Caricato, Marenich, Bloino, Janesko, Gomperts, Mennucci, Hratchian, Ortiz, Izmaylov, Sonnenberg, Williams-Young, Ding, Lipparini, Egidi, Goings, Peng, Petrone, Henderson, Ranasinghe, Zakrzewski, Gao, Rega, Zheng, Liang, Hada, Ehara, Toyota, Fukuda, Hasegawa, Ishida, Nakajima, Honda, Kitao, Nakai, Vreven, Throssell, Montgomery, Peralta, Ogliaro, Bearpark, Heyd, Brothers, Kudin, Staroverov, Keith, Kobayashi, Normand, Raghavachari, Rendell, Burant, Iyengar, Tomasi, Cossi, Millam, Klene, Adamo, Cammi, Ochterski, Martin, Morokuma, Farkas, Foresman, \& Fox}]{Frisch2016}
Frisch, M.~J., Trucks, G.~W., Schlegel, H.~B., {et~al.} 2016, Gaussian~16 {Revision C.01}

\bibitem[{{Galliano} {et~al.}(2008){Galliano}, {Madden}, {Tielens}, {Peeters}, \& {Jones}}]{Galliano2008}
{Galliano}, F., {Madden}, S.~C., {Tielens}, A. G.~G.~M., {Peeters}, E., \& {Jones}, A.~P. 2008, \apj, 679, 310

\bibitem[{{Grimme} {et~al.}(2010){Grimme}, {Antony}, {Ehrlich}, \& {Krieg}}]{Grimme2010}
{Grimme}, S., {Antony}, J., {Ehrlich}, S., \& {Krieg}, H. 2010, \jcp, 132, 154104

\bibitem[{{Grimme} {et~al.}(2011){Grimme}, {Ehrlich}, \& {Goerigk}}]{Grimme2011}
{Grimme}, S., {Ehrlich}, S., \& {Goerigk}, L. 2011, J. Comput. Chem., 32, 1456

\bibitem[{{Hanine} {et~al.}(2020){Hanine}, {Meng}, {Lu}, {Xie}, {Picaud}, {Devel}, \& {Wang}}]{Hanine2020}
{Hanine}, M., {Meng}, Z., {Lu}, S., {et~al.} 2020, \apj, 900, 188

\bibitem[{{Hudgins} {et~al.}(2005){Hudgins}, {Bauschlicher}, \& {Allamandola}}]{Hudgins2005}
{Hudgins}, D.~M., {Bauschlicher}, Jr., C.~W., \& {Allamandola}, L.~J. 2005, \apj, 632, 316

\bibitem[{{Kerkeni} {et~al.}(2022){Kerkeni}, {Garc{\'\i}a-Bernete}, {Rigopoulou}, {Tew}, {Roche}, \& {Clary}}]{Kerkeni2022}
{Kerkeni}, B., {Garc{\'\i}a-Bernete}, I., {Rigopoulou}, D., {et~al.} 2022, \mnras, 513, 3663

\bibitem[{Kroonblawd {et~al.}(2019)Kroonblawd, Lindsey, \& Goldman}]{Kroonblawd2019}
Kroonblawd, M.~P., Lindsey, R.~K., \& Goldman, N. 2019, Chem. Sci., 10, 6091

\bibitem[{{Leger} {et~al.}(1989){Leger}, {D'Hendecourt}, \& {Defourneau}}]{Leger1989}
{Leger}, A., {D'Hendecourt}, L., \& {Defourneau}, D. 1989, \aap, 216, 148

\bibitem[{{Li}(2020)}]{Li2020}
{Li}, A. 2020, Nat. Astron., 4, 339

\bibitem[{{Mai} {et~al.}(2025){Mai}, {Wang}, {Pan}, {Sch{\"o}rghuber}, {Kov{\'a}cs}, {Carrete}, \& {Madsen}}]{Mai2025}
{Mai}, X., {Wang}, Z., {Pan}, L., {et~al.} 2025, \mnras, 541, 3073

\bibitem[{{Maragkoudakis} {et~al.}(2022){Maragkoudakis}, {Boersma}, {Temi}, {Bregman}, \& {Allamandola}}]{Maragkoudakis2022}
{Maragkoudakis}, A., {Boersma}, C., {Temi}, P., {Bregman}, J.~D., \& {Allamandola}, L.~J. 2022, \apj, 931, 38

\bibitem[{{Marsh} \& {Wornat}(2000)}]{Marsh2000}
{Marsh}, N.~D. \& {Wornat}, M.~J. 2000, Proc. Combust. Inst., 28, 2585

\bibitem[{{Marshall} \& {Sadeghpour}(2016)}]{Marshall2016}
{Marshall}, D.~W. \& {Sadeghpour}, H.~R. 2016, \mnras, 455, 2889

\bibitem[{{Mattioda} {et~al.}(2020){Mattioda}, {Hudgins}, {Boersma}, {Bauschlicher}, {Ricca}, {Cami}, {Peeters}, {S{\'a}nchez de Armas}, {Puerta Saborido}, \& {Allamandola}}]{Mattioda2020}
{Mattioda}, A.~L., {Hudgins}, D.~M., {Boersma}, C., {et~al.} 2020, \apjs, 251, 22

\bibitem[{{McGuire} {et~al.}(2021){McGuire}, {Loomis}, {Burkhardt}, {Lee}, {Shingledecker}, {Charnley}, {Cooke}, {Cordiner}, {Herbst}, {Kalenskii}, {Siebert}, {Willis}, {Xue}, {Remijan}, \& {McCarthy}}]{McGuire2021}
{McGuire}, B.~A., {Loomis}, R.~A., {Burkhardt}, A.~M., {et~al.} 2021, Science, 371, 1265

\bibitem[{{Meng} \& {Wang}(2023)}]{Meng2023MD}
{Meng}, Z. \& {Wang}, Z. 2023, \mnras, 526, 3335

\bibitem[{{Meng} {et~al.}(2023){Meng}, {Zhang}, {Liang}, \& {Wang}}]{Meng2023ML}
{Meng}, Z., {Zhang}, Y., {Liang}, E., \& {Wang}, Z. 2023, \mnras, 525, L29

\bibitem[{Mermin(1968)}]{Mermin1968}
Mermin, N.~D. 1968, Phys. Rev., 176, 250

\bibitem[{{Misselt} {et~al.}(2025){Misselt}, {Witt}, {Gordon}, {Van De Putte}, {Trahin}, {Abergel}, {Noriega-Crespo}, {Guillard}, {Zannese}, {Dell'ova}, {Baes}, {Klaassen}, \& {Ysard}}]{Misselt2025A&A}
{Misselt}, K., {Witt}, A.~N., {Gordon}, K.~D., {et~al.} 2025, \aap, 700, A158

\bibitem[{{Montillaud} {et~al.}(2013){Montillaud}, {Joblin}, \& {Toublanc}}]{Montillaud2013}
{Montillaud}, J., {Joblin}, C., \& {Toublanc}, D. 2013, \aap, 552, A15

\bibitem[{{Parker} {et~al.}(2015){Parker}, {Yang}, {Dangi}, {Kaiser}, {Bera}, \& {Lee}}]{Parker2015}
{Parker}, D. S.~N., {Yang}, T., {Dangi}, B.~B., {et~al.} 2015, \apj, 815, 115

\bibitem[{{Parneix} {et~al.}(2017){Parneix}, {Gamboa}, {Falvo}, {Bonnin}, {Pino}, \& {Calvo}}]{Parneix2017}
{Parneix}, P., {Gamboa}, A., {Falvo}, C., {et~al.} 2017, Mol. Astrophys., 7, 9

\bibitem[{Patch {et~al.}(2025)Patch, McClish, Panchagnula, Rap, Banhatti, Hrodmarsson, Br{\"u}nken, Linnartz, Tielens, \& Bouwman}]{Patch2025}
Patch, M.~M., McClish, R., Panchagnula, S., {et~al.} 2025, J. Am. Chem. Soc., 147, 34508

\bibitem[{{Patra} {et~al.}(2014){Patra}, {Kr{\'a}l}, \& {Sadeghpour}}]{Patra2014}
{Patra}, N., {Kr{\'a}l}, P., \& {Sadeghpour}, H.~R. 2014, \apj, 785, 6

\bibitem[{{Peeters}(2011)}]{Peeters2011}
{Peeters}, E. 2011, in IAU Symposium, Vol. 280, The Molecular Universe, ed. J.~{Cernicharo} \& R.~{Bachiller}, 149--161

\bibitem[{{Peeters} {et~al.}(2002){Peeters}, {Hony}, {Van Kerckhoven}, {Tielens}, {Allamandola}, {Hudgins}, \& {Bauschlicher}}]{Peeters2002}
{Peeters}, E., {Hony}, S., {Van Kerckhoven}, C., {et~al.} 2002, \aap, 390, 1089

\bibitem[{Peeters {et~al.}(2021)Peeters, Mackie, Candian, \& Tielens}]{Peeters2021}
Peeters, E., Mackie, C., Candian, A., \& Tielens, A. G. G.~M. 2021, Acc. Chem. Res., 54, 1921

\bibitem[{{Peeters} {et~al.}(2004){Peeters}, {Spoon}, \& {Tielens}}]{Peeters2004}
{Peeters}, E., {Spoon}, H.~W.~W., \& {Tielens}, A.~G.~G.~M. 2004, \apj, 613, 986

\bibitem[{{Pendleton} {et~al.}(1999){Pendleton}, {Tielens}, {Tokunaga}, \& {Bernstein}}]{Pendleton1999}
{Pendleton}, Y.~J., {Tielens}, A.~G.~G.~M., {Tokunaga}, A.~T., \& {Bernstein}, M.~P. 1999, \apj, 513, 294

\bibitem[{{Rapacioli} {et~al.}(2005){Rapacioli}, {Joblin}, \& {Boissel}}]{Rapacioli2005}
{Rapacioli}, M., {Joblin}, C., \& {Boissel}, P. 2005, \aap, 429, 193

\bibitem[{{Ricca} {et~al.}(2011){Ricca}, {Bauschlicher}, \& {Allamandola}}]{Ricca2011}
{Ricca}, A., {Bauschlicher}, Jr., C.~W., \& {Allamandola}, L.~J. 2011, \apj, 729, 94

\bibitem[{{Ricca} {et~al.}(2012){Ricca}, {Bauschlicher}, {Boersma}, {Tielens}, \& {Allamandola}}]{Ricca2012}
{Ricca}, A., {Bauschlicher}, Jr., C.~W., {Boersma}, C., {Tielens}, A. G.~G.~M., \& {Allamandola}, L.~J. 2012, \apj, 754, 75

\bibitem[{{Ricca} {et~al.}(2026){Ricca}, {Boersma}, {Maragkoudakis}, {Roser}, {Shannon}, {Allamandola}, \& {Bauschlicher}}]{Ricca2026}
{Ricca}, A., {Boersma}, C., {Maragkoudakis}, A., {et~al.} 2026, \apjs, 282, 7

\bibitem[{{Rigopoulou} {et~al.}(2021){Rigopoulou}, {Barale}, {Clary}, {Shan}, {Alonso-Herrero}, {Garc{\'\i}a-Bernete}, {Hunt}, {Kerkeni}, {Pereira-Santaella}, \& {Roche}}]{Rigopoulou2021}
{Rigopoulou}, D., {Barale}, M., {Clary}, D.~C., {et~al.} 2021, \mnras, 504, 5287

\bibitem[{{Rigopoulou} {et~al.}(2024){Rigopoulou}, {Donnan}, {Garc{\'\i}a-Bernete}, {Pereira-Santaella}, {Alonso-Herrero}, {Davies}, {Hunt}, {Roche}, \& {Shimizu}}]{Rigopoulou2024}
{Rigopoulou}, D., {Donnan}, F.~R., {Garc{\'\i}a-Bernete}, I., {et~al.} 2024, \mnras, 532, 1598

\bibitem[{{Robitaille} {et~al.}(2020){Robitaille}, {Deil}, \& {Ginsburg}}]{Robitaille2020}
{Robitaille}, T., {Deil}, C., \& {Ginsburg}, A. 2020, {reproject: Python-based astronomical image reprojection}, Astrophysics Source Code Library, record ascl:2011.023

\bibitem[{{Sellgren} {et~al.}(2010){Sellgren}, {Werner}, {Ingalls}, {Smith}, {Carleton}, \& {Joblin}}]{Sellgren2010}
{Sellgren}, K., {Werner}, M.~W., {Ingalls}, J.~G., {et~al.} 2010, \apjl, 722, L54

\bibitem[{{Singh} {et~al.}(2025){Singh}, {Biennier}, {Simon}, {Chakraborty}, {Dartois}, {Georges}, {Kassi}, {Chandrasekaran}, {Sabbah}, {Joblin}, {Ranjan}, {Suwas}, {Gopalan}, \& {Arunan}}]{Singh2025}
{Singh}, D.~K., {Biennier}, L., {Simon}, A., {et~al.} 2025, \aap, 704, A345

\bibitem[{{Skretas} {et~al.}(2025){Skretas}, {Karska}, {Francis}, {Rocha}, {van Gelder}, {Tychoniec}, {Figueira}, {Sewi{\l}o}, {Wyrowski}, \& {Schilke}}]{Skretas2025}
{Skretas}, I.~M., {Karska}, A., {Francis}, L., {et~al.} 2025, \aap, 703, A139

\bibitem[{{Smith} {et~al.}(2007){Smith}, {Draine}, {Dale}, {Moustakas}, {Kennicutt}, {Helou}, {Armus}, {Roussel}, {Sheth}, {Bendo}, {BucKalew}, {Calzetti}, {Engelbracht}, {Gordon}, {Hollenbach}, {Li}, {Malhotra}, {Murphy}, \& {Walter}}]{Smith2007}
{Smith}, J.~D.~T., {Draine}, B.~T., {Dale}, D.~A., {et~al.} 2007, \apj, 656, 770

\bibitem[{{Spilker} {et~al.}(2023){Spilker}, {Phadke}, {Aravena}, {Archipley}, {Bayliss}, {Birkin}, {B{\'e}thermin}, {Burgoyne}, {Cathey}, {Chapman}, {Dahle}, {Gonzalez}, {Gururajan}, {Hayward}, {Hezaveh}, {Hill}, {Hutchison}, {Kim}, {Kim}, {Law}, {Legin}, {Malkan}, {Marrone}, {Murphy}, {Narayanan}, {Navarre}, {Olivier}, {Rich}, {Rigby}, {Reuter}, {Rhoads}, {Sharon}, {Smith}, {Solimano}, {Sulzenauer}, {Vieira}, {Vizgan}, {Wei{\ss}}, \& {Whitaker}}]{Spilker2023}
{Spilker}, J.~S., {Phadke}, K.~A., {Aravena}, M., {et~al.} 2023, \nat, 618, 708

\bibitem[{{Stuart} {et~al.}(2000){Stuart}, {Tutein}, \& {Harrison}}]{Stuart2000}
{Stuart}, S.~J., {Tutein}, A.~B., \& {Harrison}, J.~A. 2000, \jcp, 112, 6472

\bibitem[{{Tappe} {et~al.}(2012){Tappe}, {Rho}, {Boersma}, \& {Micelotta}}]{Tappe2012}
{Tappe}, A., {Rho}, J., {Boersma}, C., \& {Micelotta}, E.~R. 2012, \apj, 754, 132

\bibitem[{{Tielens}(2008)}]{Tielens2008}
{Tielens}, A.~G.~G.~M. 2008, \araa, 46, 289

\bibitem[{{Tsuge} {et~al.}(2023){Tsuge}, {Molpeceres}, {Aikawa}, \& {Watanabe}}]{Tsuge2023}
{Tsuge}, M., {Molpeceres}, G., {Aikawa}, Y., \& {Watanabe}, N. 2023, Nat. Astron., 7, 1351

\bibitem[{{Van De Putte} {et~al.}(2026){Van De Putte}, {Gordon}, {Misselt}, {Witt}, {Abergel}, {Noriega-Crespo}, {Guillard}, {Zannese}, {Elyajouri}, {Trahin}, {Dell'ova}, {Baes}, \& {Klaassen}}]{VanDePutte2026}
{Van De Putte}, D., {Gordon}, K.~D., {Misselt}, K., {et~al.} 2026, \aap, 710, A60

\bibitem[{{van Dishoeck}(2004)}]{vanDishoeck2004}
{van Dishoeck}, E.~F. 2004, \araa, 42, 119

\bibitem[{{van Duin} {et~al.}(2001){van Duin}, {Dasgupta}, {Lorant}, \& {Goddard}}]{Duin2001}
{van Duin}, A. C.~T., {Dasgupta}, S., {Lorant}, F., \& {Goddard}, W.~A. 2001, J. Phys. Chem. A, 105, 9396

\bibitem[{{Vats} {et~al.}(2022){Vats}, {Pathak}, {Onaka}, {Buragohain}, {Sakon}, \& {Endo}}]{Vats2022}
{Vats}, A., {Pathak}, A., {Onaka}, T., {et~al.} 2022, \pasj, 74, 161

\bibitem[{{Wang}(2026)}]{Wang2026}
{Wang}, Z. 2026, \aap, 710, L17

\bibitem[{{Weippert} {et~al.}(2020){Weippert}, {Hauns}, {Bachmann}, {Greisch}, {Narita}, {M{\"u}llen}, {B{\"o}ttcher}, \& {Kappes}}]{Weippert2020}
{Weippert}, J., {Hauns}, J., {Bachmann}, J., {et~al.} 2020, J. Phys. Chem. C, 124, 8236

\bibitem[{{Werner} {et~al.}(2004){Werner}, {Uchida}, {Sellgren}, {Marengo}, {Gordon}, {Morris}, {Houck}, \& {Stansberry}}]{Werner2004}
{Werner}, M.~W., {Uchida}, K.~I., {Sellgren}, K., {et~al.} 2004, \apjs, 154, 309

\bibitem[{{Yang} {et~al.}(2017){Yang}, {Li}, {Glaser}, \& {Zhong}}]{Yang2017}
{Yang}, X.~J., {Li}, A., {Glaser}, R., \& {Zhong}, J.~X. 2017, \apj, 837, 171

\bibitem[{{Yu} \& {Nyman}(2012)}]{Yu2012}
{Yu}, H.-G. \& {Nyman}, G. 2012, \apj, 751, 3

\end{thebibliography}

\FloatBarrier
\clearpage

\end{document}